\documentclass[usenatbib]{mnras}
\usepackage[dvips]{graphicx}
\usepackage[english]{babel}
\usepackage{amsmath}
\usepackage{amssymb}
\usepackage{newtxtext,newtxmath}
\usepackage{float}

\newcommand{\EQ}{\begin{equation}}
\newcommand{\EN}{\end{equation}}
\newcommand{\EQA}{\begin{eqnarray}}
\newcommand{\ENA}{\end{eqnarray}}

\newcommand{\bfl}{\mbox{\boldmath $l$} {}}
\newcommand{\kk}{\mbox{\boldmath $k$} {}}
\newcommand{\BB}{\mbox{\boldmath $B$} {}}

\newcommand{\mkG}{\,{\rm \mu G}}

\newcommand{\sigmarm}{\bar{\sigma}_{\rm RM}}

\def\Rm{\rm Rm}

\def\Rey{{\rm Re}}
\def\Pm{\rm Pm}
\title[Decaying turbulence and magnetic fields]{Decaying turbulence and magnetic fields in galaxy clusters}
\author[Sur]{Sharanya Sur$^{1}$\thanks{E-mail: sharanya.sur@iiap.res.in} \\
$^1$Indian Institute of Astrophysics, 2nd Block, Koramangala, Bangalore 560034, India}

\begin{document}

\pagerange{\pageref{firstpage}--\pageref{lastpage}} \pubyear{2019}

\maketitle

\label{firstpage}

\begin{abstract}
We explore the decay of turbulence and magnetic fields generated by fluctuation dynamo action 
in the context of galaxy clusters where such a decaying phase can occur in the aftermath of a 
major merger event. Using idealized numerical simulations that start from a kinetically dominated 
regime we focus on the decay of the steady state rms velocity and the magnetic field for a wide 
range of conditions that include varying the compressibility of the flow, the forcing wave number,  
and the magnetic Prandtl number. Irrespective of the compressibility of the flow, both the rms 
velocity and the rms magnetic field decay as a power-law in time. In the subsonic case we find 
that the exponent of the power-law is consistent with the $-3/5$ scaling reported in previous 
studies. However, in the transonic regime both the rms velocity and the magnetic field initially 
undergo rapid decay with an $\approx t^{-1.1}$ scaling with time. This is followed by a phase of 
slow decay where the decay of the rms velocity exhibits an $\approx -3/5$ scaling in time, while 
the rms magnetic field scales as $\approx -5/7$. Furthermore, analysis of the Faraday rotation 
measure reveals that the Faraday RM decays also decays as a power law in time $\approx t^{-5/7}$; 
steeper than the $\sim t^{-2/5}$ scaling obtained in previous simulations of magnetic field decay 
in subsonic turbulence. Apart from galaxy clusters, our work can have potential implications in the 
study of magnetic fields in elliptical galaxies. 
\end{abstract}


\section{Introduction}

Free decay of magnetohydrodynamic (MHD) turbulence has been an active area of research 
in a variety of contexts. Starting from the generation of large-scale primordial magnetic fields in 
the early Universe \citep{CHB01, BJ04, SS05, BKT15, B+17b, RB17} decaying MHD turbulence 
has also been studied in connection with the recent detection of strong optical polarization in 
gamma-ray burst (GRB) afterglows \citep{Uehara+12, Mundell+13, Z14} as well as to measure 
the decay rates of sub and super-Alfv\'{e}nic supersonic turbulence in the interstellar medium 
\citep[ISM;][]{Maclow+98}. Work in primordial magnetic fields focussed on the occurrence of inverse 
cascade of magnetic energy from small to large-scales in presence or absence of magnetic 
helicity and how the cascade depends on the magnetic Prandtl number ($\Pm$) and Reynolds 
numbers. In the absence of helicity \citet{BJ04} showed that the decay law of magnetic 
energy ($E_{\rm m}$) depends on the large-scale spectral index ($n$) of magnetic field fluctuations. 
Specifically, $E_{\rm m}(t) \sim t^{-6/5}$ for $n=2$ and $\sim t^{-10/7}$ when one assumes a blue 
spectrum (i.e. $n=4$) for the field. Remarkably, these power law scalings are similar to what one 
obtains in the decay of kinetic energy in isotropic hydrodynamic turbulence where the exponent of 
the power law depends on the scaling of the energy spectrum $E(k)$ at low $k$. For instance, if 
$E (k \rightarrow 0) \sim k^{4}$, the turbulent kinetic energy $u^{2}(t) \sim t^{-10/7}$ and for 
$E (k\rightarrow 0) \sim k^{2}, u^{2}(t) \sim t^{-6/5}$ 
\citep[][and references therein]{BP56, Saff67, LO00, IDK06, DOK12}. More recently \citet{BKT15} 
reported an inverse cascade of energy with $E_{\rm m}(t) \propto t^{-1}$ even in the absence of 
helicity. This was later confirmed by \citet{Z14} but the cascade is likely to be suppressed at large 
$\Pm$ \citep{RB17}. 

A common feature in the above mentioned works on primordial magnetic fields and the work of  
\citet{Z14} is that they all start from a magnetically dominated regime with the field initialized by
a power spectrum peaked at around a large wavenumber. In galaxies and clusters the magnetic 
Reynolds numbers ($\Rm$) are large enough to excite a Fluctuation/small-scale dynamo 
\citep{K68, ZRS90} which amplifies dynamically negligible seed fields by turbulent stretching of the 
field lines. The resulting field exhibits an intermittent structure with the magnetic energy spectrum 
peaked at resistive scales at early times which then gradually shifts to larger scales (i.e. to smaller 
$k$) as the dynamo approaches saturation. How does such a configuration decay? 
Note that a key difference here is that in contrast to a magnetically dominated regime one starts 
in the kinetically dominated regime, grows and saturates the field to a fraction of the equipartition 
value and then allows for the free decay of turbulence along with the field. Apart from the obvious 
excitement about gaining theoretical insights of such a process, free decay of magnetic fields 
initially amplified by fluctuation dynamo is of relevance in galaxy clusters in the aftermath of 
epochs of major mergers. 
Although turbulence and magnetic fields in galaxy clusters have been studied in great detail 
\citep{DBL99,DBL02,Bru+05,Ryu+08,CR09, BS13,PJR15,Mari+15,Vazza+17, Vazza+18, 
Mari+18,Dom+19,MS19,Rho+19,SZ19}, one of the early works which explored the decay of dynamo 
generated fields in these systems is by \citet{SSH06}. Aided by incompressible non-helical turbulence 
simulations, they observed that both the field and turbulent speed undergo a power-law decay, 
decreasing by a factor of 2 during this stage, whereas their scales increase by about the same factor. 
While central cluster volumes are expected to be subsonic (and dominated by solenoidal 
modes), transonic turbulence can occur in the outskirts \citep{Ryu+08, Paul+11, Min14, Min15, 
Vazza+17, Donnert+18}. 
In this work, we revisit and expand the study with the aim to probe how non-helical turbulence and 
magnetic fields generated by fluctuation dynamos decay under different conditions such as the 
compressibility of the flow, forcing scale of turbulent driving, and varying magnetic Prandtl numbers. 

The content of this paper is organized as follows. In Section \ref{sims} we outline the numerical 
set-up, initial conditions and parameter space covered by our simulations. Thereafter, in section 
\ref{results}, we present the results spread over two subsections. In the first, we focus on the decay 
of subsonic turbulence which was driven purely by solenoidal modes. In particular, we discuss the 
time evolution of the field structure, rms velocity and rms magnetic field and the power spectra 
obtained from the different runs. In the next subsection, we describe the results obtained from a 
simulation where turbulence was driven by a mixture of solenoidal and compressive modes with 
the amplitude of the forcing adjusted such that the steady state velocity was transonic. In 
Section~\ref{frm}, we analyze the temporal evolution of the Faraday rotation measure (RM) in the 
case of decaying transonic turbulence. Finally, in Section~\ref{conc}, we summarize the main results 
and discuss future research directions emanating from this work. 

\section{Numerical set-up and initial conditions} \label{sims}

In order to study the dynamics of decaying non-helical turbulence and magnetic fields starting 
from a kinetically dominated regime, we performed a suite of non-ideal, three-dimensional 
simulations of fluctuation dynamos with the {\scriptsize FLASH} code \citep{Fryxell+00}. In this 
code, shocks that naturally occur in transonic and supersonic turbulence are accurately handled 
by Riemann solvers without using artificial viscosity. The numerical setup used to perform 
these non-ideal compressible simulations is identical to the one described in \citet{SBS18}. 
Specifically, we adopt an isothermal equation of state with initial density and sound speed set to 
unity in a box of unit length having $256^{3} - 512^{3}$ grid points with periodic boundary conditions. 
Turbulence is driven as a stochastic Ornstein--Uhlenbeck process \citep{EP88, Benzi+08}, with a 
finite time correlation. Recent studies of turbulence in galaxy clusters suggest that the kinetic energy 
of cluster turbulence is dominated by the solenoidal component \citep{Min15, Vazza+17}. While the 
dominance of such modes is expected in cluster cores, compressive motions are also likely to 
contribute in the outskirts. We therefore, chose to drive turbulence using purely solenoidal modes 
for the subsonic simulations and used a mixed forcing prescription as described in \citet{Fed+10}
to drive turbulence in the transonic case. More details about the latter are presented in the 
subsection~\ref{decay_trans}. In the runs presented here, we have chosen the forcing scale of driven 
turbulence to be either in the range $1 \leq |\kk| L/2\pi \leq 3$ such that the average forcing 
wavenumber $k_{\rm f}L/2\pi \sim 2$ or at larger $k$ in the range $4 \leq |\kk| L/2\pi \leq 6$ 
($k_{\rm f}L/2\pi \sim 5$). 
Here $L$ is the length of the box. The former corresponds to forcing near the box scale thereby 
containing only a few turbulent cells, while the latter corresponds to driving motions on very small 
scales resulting in a larger number of turbulent cells. We also adjust the strength of the forcing to 
achieve a steady state rms Mach number $\mathcal{M} \approx 0.15 - 0.18$ in the subsonic runs 
and $\mathcal{M} \approx 1$ in the transonic case. The initial magnetic field was chosen to be of the 
form $\BB = B_{0} [0,0,\sin(10\pi x)]$ with the amplitude $B_{0}$ adjusted to a value such that the 
initial plasma $\beta \sim 10^{6}$. Further, to maintain divergence of the magnetic field to machine 
precision level, we use the unsplit staggered mesh algorithm in {\scriptsize FLASH} with a constrained 
transport scheme \citep{LD09, Lee13} and Harten-Lax-van Leer-Discontinuities (HLLD) Riemann 
solver \citep{MK05} to resolve shocks.

%%%%%%%%%%%%%%%%%%%%%%%%%%%%%%%%%%%%%%%%%%%%%%%%%%%%
\begin{table}
\setlength{\tabcolsep}{8.0 pt}
\caption{Key parameters for various simulation runs. 
$\mathcal{M}$ is the average value of the rms Mach number obtained in the 
steady state before the forcing is turned off. In the last column $t_{\rm 0}$ 
refers to the time (in code units) when the turbulent forcing was turned off. 
}
\begin{tabular}{|c|c|c|c|c|c|c|}
\hline
\hline
Run & $N^{3}$ & $k_{\rm f}$ & $\mathcal{M}$ & $\Pm$ & $\Rey = u\,l_{\rm f}/\nu$ & $t_{0}$ \\
\hline
A & $512^{3}$ & 2.0 & 0.18 & 1 & 1080 & 80.6 \\
B & $512^{3}$ & 2.0 & 0.17 & 5 & 350 & 61.0 \\
C & $256^{3}$ & 5.0 & 0.17 & 1 & 420 & 62.6 \\
D & $512^{3}$ & 5.0 & 0.15 & 5 & 250 & 25.2 \\
E & $512^{3}$ & 5.0 & 1.0 & 1 & 1200 & 6.55 \\ 
\hline
\hline
\label{sumsim}
\end{tabular}
\end{table}
%%%%%%%%%%%%%%%%%%%%%%%%%%%%%%%%%%%%%%%%%%%%%%%%%%%%

Table~\ref{sumsim} provides a summary of the parameters of the runs considered here. In order to simulate 
forced and then decaying turbulence, the flow is driven until the nonlinear saturated phase of the dynamo 
has been captured over sufficient eddy turnover times defined as $t_{\rm ed} = l_{\rm f}/u_{\rm rms}$, after 
which the forcing is switched off. Here $l_{\rm f} = 2\pi/k_{\rm f}$ is the forcing scale and time $t_{\rm 0}$ in 
Table~\ref{sumsim} refers to the time in code units when the driving is halted. In what follows, we start by 
discussing the results obtained in the decaying regime when turbulence was forced subsonically. 
%%%%%%%%%%%%%%%%%%%%%%%%%%%%%%%%%%%%%%%%%%%%%%%%%%%
\begin{figure*}
\includegraphics[width=\textwidth]{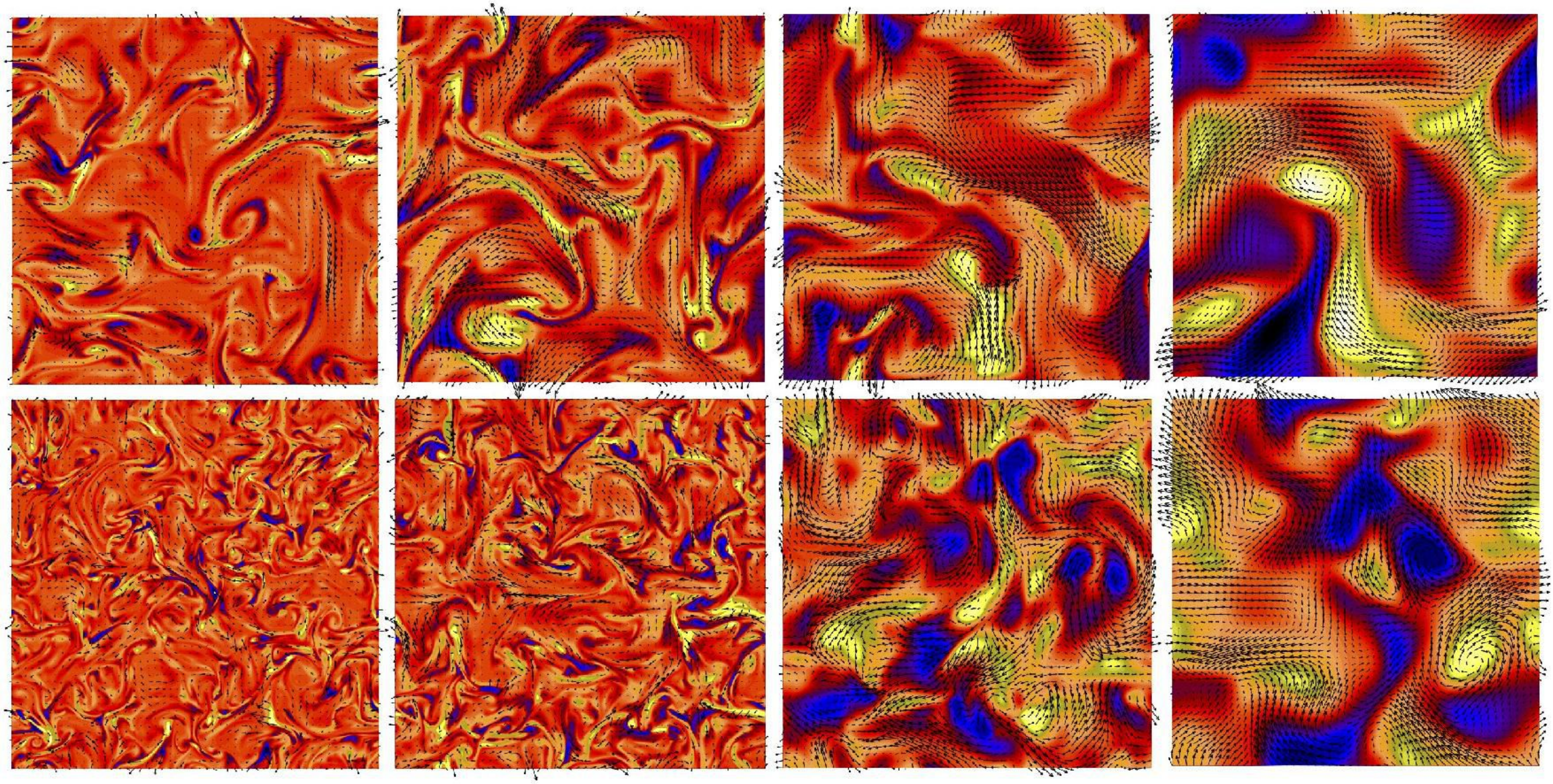}
\caption{Two-dimensional slices of $B_{z}/B_{\rm rms}$ in the $x-y$ plane from runs B (top row) 
and D (bottom row), showing the field structure in kinematic (first column), saturated (second 
column) and decaying phases (last two columns). The fields are strong in blue and yellow regions 
and negligible in orange regions. The field in the plane of the slices are shown with vectors whose 
length is proportional to the field strength. For clarity of the color contrast, we have restricted the 
range of $B_{z}/B_{\rm rms}$ to be $\pm3.0$ in the kinematic and saturated phase and to
$\pm1.5$ in the decay phases. 
}
\label{evolmag}
\vspace{-1.5em}
\end{figure*}
%%%%%%%%%%%%%%%%%%%%%%%%%%%%%%%%%%%%%%%%%%%%%%%%%%%
\section{Results}
\label{results}

\subsection{Decaying regime in subsonic turbulence}
\label{decay_sub}

\subsubsection{Field structure and time evolution}
\label{timevol}

%%%%%%%%%%%%%%%%%%%%%%%%%%%%%%%%%%%%%%%%%%%%%%%%%%%
\begin{figure}
\begin{center}
\includegraphics[width=1.05\columnwidth]{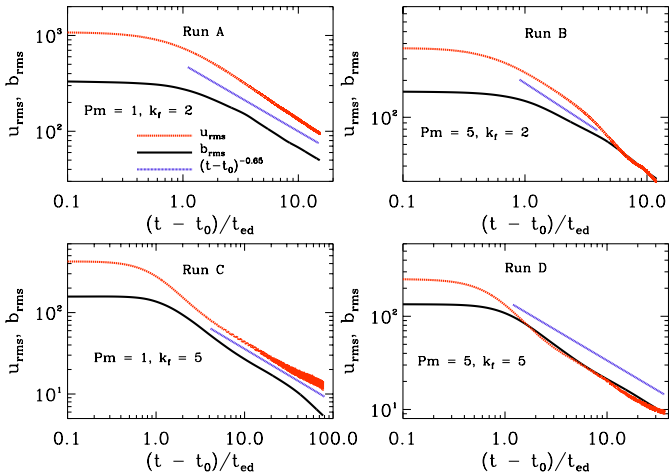}
\vspace{-1em}
\caption{Time evolution of $u_{\rm rms}$ and $b_{\rm rms}$ in the decaying regime as a function 
of the eddy turnover time. Both variables are normalized in units of ($l_{\rm f}/\nu$), while time is 
expressed in units of $t' = (t - t_{0})/t_{\rm ed}$. The blue dotted lines in each panel shows the 
$t'^{(-0.65)}$ scaling.
\label{tevol_decay}}
\end{center}
\vspace{-1.5em}
\end{figure}
%%%%%%%%%%%%%%%%%%%%%%%%%%%%%%%%%%%%%%%%%%%%%%%%%%%%%%

In Fig.~\ref{evolmag}, we show the two-dimensional structure of the magnetic field in kinematic 
(first column), saturated (second column), and decaying phases (last two columns) for runs B and D. 
Both these runs have $\Pm = 5$. However the forcing in run B is at $k_{\rm f}=2$, while it is at 
$k_{\rm f} = 5$ in run D. The slices are in the $x-y$ plane and the $z$-component of the magnetic 
field normalized to its rms value is shown in color. The field in the plane of the slices are shown with 
vectors whose length is proportional to the field strength. Because of the presence of only a few 
turbulent cells, stretching of the magnetic field lines by turbulent eddies is more prominently visible 
in run B compared to run D (see second column). Field structure at this stage is expected to be highly 
intermittent having complex internal structure. Once the turbulent forcing is turned off, the intermittency 
gradually decreases as the field decays together with the turbulence. Structures on smaller scales 
decay faster and the volume filling factor of the field increases with time. This results in the field 
structure becoming more homogeneous in both the runs. The magnetic field structures in $\Pm = 1$ 
runs (not shown here) also show qualitatively similar features. 
%%%%%%%%%%%%%%%%%%%%%%%%%%%%%%%%%%%%%%%%%%%%%%%%%%%%%%
\begin{figure*}
\includegraphics[width=\textwidth]{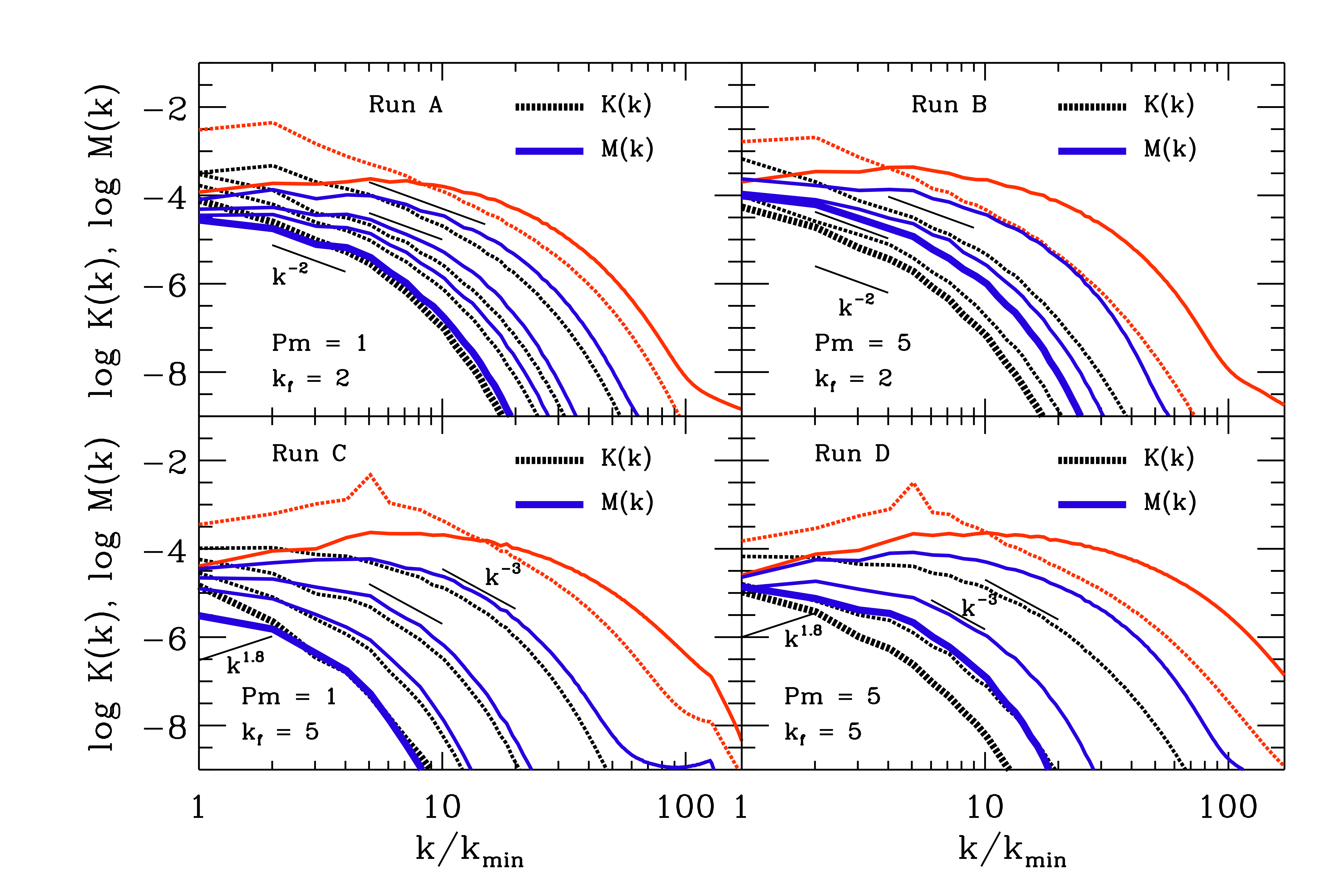}
\caption{The spectra of kinetic (dotted) and magnetic (solid) energies at various stages of evolution,  
starting from the time turbulent forcing is turned off. In each panel, kinetic and magnetic energies at 
that time is shown as red dotted and solid lines respectively. The final kinetic and magnetic energy 
spectra in the decay phase are plotted as thick dotted and thick blue solid lines, respectively. 
Wavenumber is normalized in units of $k_{\rm min} = 2\pi$.}
\label{spec_sub}
\vspace{-1.5em}
\end{figure*}
%%%%%%%%%%%%%%%%%%%%%%%%%%%%%%%%%%%%%%%%%%%%%%%%%
 
The time evolution of the fluctuation dynamo generated magnetic field in the kinematic stage 
and all the way up to saturation has been thoroughly studied by a number of authors. Starting from 
a weak initial seed field, random stretching and folding of the field lines by the turbulent eddies leads 
to an exponential growth of the magnetic energy eventually saturating at about $10 - 30$ per cent of 
the equipartition value \citep[e.g.][]{HBD04,Scheko+04,BS05,Cho+09,Fed+11,BS13,SBS18}. However 
as our main objective is to explore the decay of the turbulent velocity and the magnetic field, we show 
in Fig.~\ref{tevol_decay} the time evolution of the rms velocity ($u_{\rm rms}$) by red dashed lines 
and the magnetic field ($b_{\rm rms}$) by black solid lines, after the driving was switched off.
For clarity in accurately estimating the power-law slopes, we choose to multiply the $u_{\rm rms}$ and 
$b_{\rm rms}$ by $(l_{\rm f}/\nu)$ while time in the abscissa is represented in units of 
$t' = (t - t_{0})/t_{\rm ed}$, where we use the steady state value of $u_{\rm rms}$ in the estimate of 
$t_{\rm ed}$. The figure shows that in all the runs both $u_{\rm rms}$ and $b_{\rm rms}$ continue 
to remain approximately constant for about an eddy-turnover time before eventual decay. The decay of 
both $u_{\rm rms}$ and $b_{\rm rms}$ can be approximated by $(t - t_{0})^{-0.65}$ power-law scaling 
(blue dotted line), which is in close agreement with the $-3/5$ scaling obtained from simple analytical 
estimates \citep[see section 2.2 of][]{SSH06}. The decay of $u_{\rm rms}$ slows down once the velocity 
reaches the box scale. Note that to achieve $\Pm > 1$, we have increased the viscosity thereby lowering 
the Reynolds number. Because of this runs B and D show that the initial decay of $u_{\rm rms}$ is 
faster than that of $b_{\rm rms}$ before slowing down. In a nutshell, we find that in subsonic turbulence 
the decay of the rms velocity and the magnetic field is in close agreement with the $-3/5$ power-law 
scaling and independent of the forcing wave numbers and magnetic Prandtl numbers explored here.

\subsubsection{Power Spectra}

The evolution of kinetic $K(k)$ and magnetic $M(k)$ energy spectra in the kinematic growth phase 
and saturated phase of fluctuation dynamos is also known from earlier studies. We therefore
focus our discussion only for the decay regime. Fig.~\ref{spec_sub} shows the temporal evolution of 
$K(k)$ and $M(k)$ denoted by dotted and solid lines, respectively. To guide the readers, we begin by 
plotting $K(k)$ and $M(k)$ at a time just before the turbulent forcing is turned off. By then the magnetic 
field is already saturated. These are denoted by thin dotted red and solid red lines for $K(k)$ and $M(k)$, 
respectively. Note that in agreement with previous studies \citep{SSH06,BS13,PJR15,SBS18}, $M(k)$ 
at saturation exceeds $K(k)$ on all but the largest scales. After the forcing is turned off the velocity 
can still drive turbulent motions on scales where the kinetic energy exceeds the magnetic energy. 
As shown in Fig.~\ref{tevol_decay} these motions can sustain dynamo action for about an eddy turnover 
time before terminal decay of turbulence and field sets in.  

The overall picture that emerges from Fig.~\ref{spec_sub} is the following. Both $K(k)$ and $M(k)$ 
collapses from the high-$k$ end in all the runs. In this phase, run A shows a $\sim k^{-2}$ 
scaling of both $K(k)$ and $M(k)$ in the range $5 < k < 15$. Gradually the peak of $K(k)$ shifts to 
the box scale beyond which the decay is expected to slow down. Similarly $M(k)$ which was peaked 
around $k \sim 7-9$ in the steady state also approaches the box scale. This implies that the field 
distribution gradually becomes more homogeneous in agreement with Fig.~\ref{evolmag}. The spectra 
at the final time is shown by thick dotted line for $K(k)$ and by thick solid blue line for $M(k)$ in all the 
panels. Similar behavior is obtained in run B ($\Pm = 5$) where both $K(k)$ and $M(k)$ also show a 
$\sim k^{-2}$ scaling during the initial phase of the decay ultimately reaching the box scale. In runs C 
and D, turbulence was forced at $k_{\rm f} = 5$. In the saturated state $M(k)$ peaks at $k \approx 10$ 
for run C and between $k = 10$ and $15$ in run D. In both runs, the spectra shows that at the high-$k$ 
end both $K(k)$ and $M(k)$ scale as $k^{-3}$ in a narrow range of wave numbers. 
On the other hand, at the low-$k$ end, $M(k)$ shows a $\sim k^{1.8}$ scaling in both runs during the 
initial decay which then flattens once the magnetic field is homogenized on the scale of the box. 

\subsection{Decaying regime in transonic turbulence}
\label{decay_trans} 

We have so far considered the free decay of MHD turbulence in subsonic flows that were 
artificially driven only by solenoidal modes. Does the decay behavior change when turbulent 
driving results in transonic flows that are likely to occur in cluster outskirts?
Before delving into the results, we first discuss the applicability of driving turbulence purely 
by solenoidal modes in these regions. Recent cosmological simulations of hierarchical structure 
formation by \citet{Min15} and \citet{Vazza+17} show that about $50-80$ per cent of the kinetic 
energy in cluster turbulence is still dominated by solenoidal modes with the contribution from the 
compressional modes only increasing during merger events. \citet{Vazza+17} further show that 
while solenoidal turbulence dominates the dissipation of turbulent motions in the central cluster 
volume at all epochs, dissipation via compressive modes is important at large radii and close to 
merger events. In the context of our numerical set-up, the above findings suggest that instead of 
driving turbulence only by solenoidal modes, using a combination of sheared (solenoidal) and 
compressive forcing is likely to be more closer to reality. 
To this effect, we make use of the mixed forcing implementation detailed in \citet{Fed+10, Fed+11} 
and report the results obtained from a $512^{3}$ simulation (run E) with $\Rey = \Rm \approx 1200$, 
where the amplitude of the forcing was adjusted such that the steady state $\mathcal{M} \approx 1$. 
Although the outer correlation scales of gas motions in cluster outskirts are expected to be of the 
order of the curvature radius of the shocks \citep[which is of order Mpc;][]{Ryu+08}, the eddy 
turnover time on the outer scale will be larger than on smaller scales. We therefore chose to force 
turbulence at $k_{\rm f} = 5$. Keeping all other conditions the same (i.e. the basic setup, the initial 
and boundary conditions as described in Sec.~\ref{sims}), we use the mixed forcing implementation 
of \citet{Fed+10} where the force field is decomposed into its solenoidal and compressive parts by 
applying a projection in Fourier space. In index notation, the projection operator can be expressed
as $\mathcal{P}^{\zeta}_{ij}({\kk}) = \zeta\,\mathcal{P}^{\perp}_{ij} + (1 - \zeta)\,\mathcal{P}^{\parallel}_{ij}$
where $\mathcal{P}^{\perp}_{ij}$ and $\mathcal{P}^{\parallel}_{ij}$ are the solenoidal and compressive 
projection operators, respectively, and the parameter $\zeta = [0,1]$\footnote{Setting $\zeta = 0$ 
results in purely compressive forcing while using $\zeta = 1$ results in purely solenoidal forcing.}. 
Choosing $\zeta = 0.2$, we find that at $\mathcal{M} \approx 1$ this results in a solenoidal ratio, 
i.e. the ratio of the specific kinetic energy in solenoidal modes to the total specific kinetic energy, 
$\chi_{\rm sol} = e_{\rm sol}/e_{\rm tot} \approx 0.8$, in the saturated phase of the dynamo. 
This could be due to vorticity generation in interacting shock waves \citep{ST03,Kritsuk+07,Ryu+08, PJR15}. 

%%%%%%%%%%%%%%%%%%%%%%%%%%%%%%%%%%%%%%%%%%%%%%%%%%%%%%%%
\begin{figure}
\begin{center}
\includegraphics[width=1.05\columnwidth]{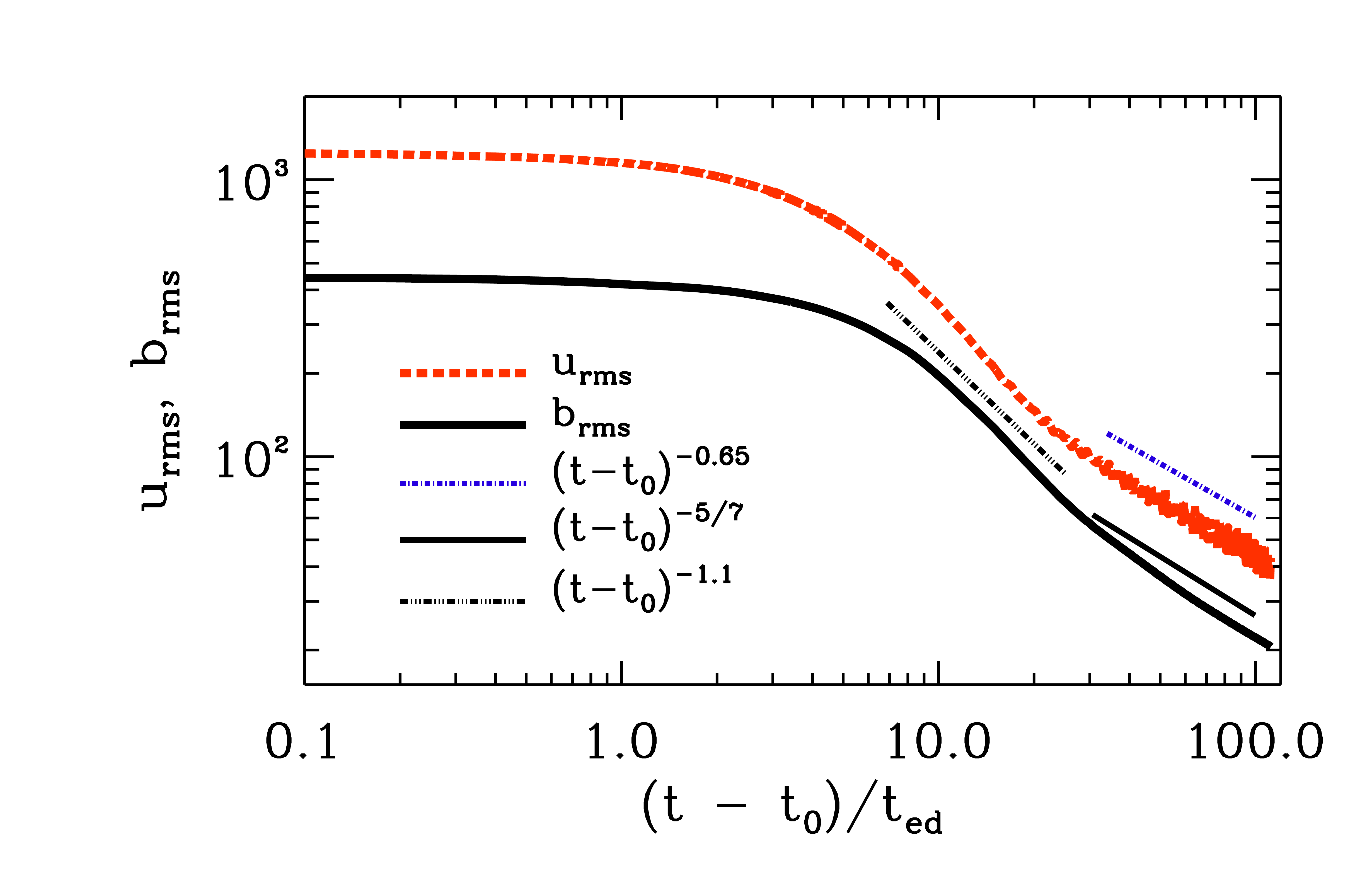}
\vspace{-1em}
\caption{Same as in Fig.~\ref{tevol_decay}, but now for run E with $\mathcal{M} \approx 1, {\rm Pm} = 1$. 
\label{tevol_trans_flash}}
\end{center}
\vspace{-1.5em}
\end{figure}
%%%%%%%%%%%%%%%%%%%%%%%%%%%%%%%%%%%%%%%%%%%%%%%%%%%%%%%
%%%%%%%%%%%%%%%%%%%%%%%%%%%%%%%%%%%%%%%%%%%%%%%%%%%%%%%
\begin{figure}
\begin{center}
\includegraphics[width=1.05\columnwidth]{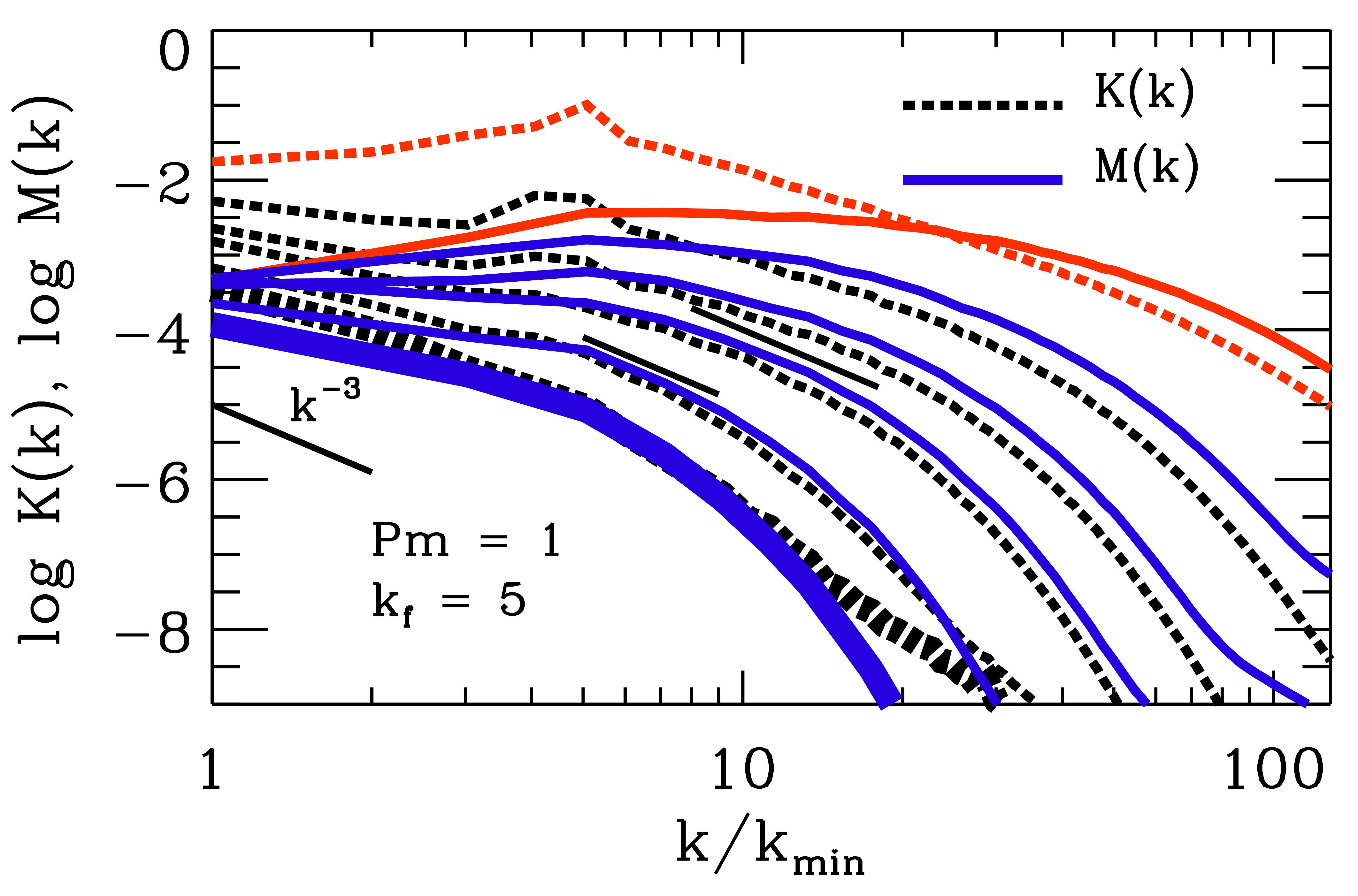}
\vspace{-1em}
\caption{Evolution of $K(k)$ and $M(k)$ in the decay phase for run E starting from the steady state (thin red 
dashed and red solid lines). Thin black dashed and thin blue solid lines are shown at times $t' \approx 10, 18,
27$ and $55$. The final $K(k)$ and $M(k)$ spectra at $t'=112$ are shown by thick dashed black and thick 
blue solid lines, respectively. Wavenumber is normalized in units of $k_{\rm min} = 2\pi$. 
\label{spec_trans}}
\end{center}
\end{figure}
%%%%%%%%%%%%%%%%%%%%%%%%%%%%%%%%%%%%%%%%%%%%%%%%%%%%%%%
Figure~\ref{tevol_trans_flash} shows the evolution of the normalized $u_{\rm rms}$ and $b_{\rm rms}$ 
from the transonic run. Both the velocity and the magnetic field continue to evolve in steady state for 
$\approx 2-3\,t_{\rm ed}$ after turbulent forcing has been switched off. Subsequently, during the interval 
$t' = 7 - 20$, both $u_{\rm rms}$ and $b_{\rm rms}$ rapidly decays approximately as $t'^{(-1.1)}$. 
A possible cause for the brisk decrease of $u_{\rm rms}$ could be due to the decay of shocks and 
shock fronts. Beyond this phase, a change in slope is observed with $u_{\rm rms}$ scaling approximately 
as $t'^{(-0.65)}$ and $b_{\rm rms}$ scaling $\approx t'^{(-5/7)}$. The reason for the $t'^{(-0.65)}$ scaling 
of $u_{\rm rms}$ can be understood from the fact that once turbulent driving is turned off, the rms 
$\mathcal{M}$ also starts to decay. In fact, we find that by $t' \sim 20\,t_{\rm ed}$ the rms Mach number 
decreased by a factor of ten from the steady state value $\mathcal{M} \approx 1$. What follows thereafter 
is the terminal decay of subsonic motions. Fig.~\ref{spec_trans} shows time evolution of $K(k)$ and $M(k)$ 
in the decay phase. As before, the red dashed and red solid lines are the spectra in the non-linear stage 
just before the turbulent forcing is switched off. Subsequently thin black dashed and blue solid lines depict 
the spectra of $K(k)$ and $M(k)$, respectively at times $t' = 10, 18, 27$ and $55$. Early into the decay, 
$K(k)$ has a $\sim k^{-3}$ slope in the range $k \sim 8 - 20$ and thereafter in the range $k \sim 3 - 6$. 
The final spectra at $t' = 112$ are denoted by thick lines.

\section{Faraday rotation measure in decaying turbulence}
\label{frm}

An important observational diagnostic of intracluster magnetic fields is the Faraday rotation 
of polarized radio emission of background radio sources as seen through the cluster \citep{CKB01, 
B+10, B+13, Govoni+10, Govoni+17, Han17}. This is quantified by the Faraday rotation measure (RM) 
defined as, ${\rm RM} = K\int_{L}\,n_{\rm e}\BB\cdot d{\bfl}$. 
Here $K = 0.81\,{\rm rad\,m^{-2}\,cm^{3}\,\mkG^{-1}\,pc^{-1}}$ is a constant, $n_{\rm e}$ is the 
thermal electron density measured in ${\rm cm^{-3}}$, $\BB$ is the magnetic field in $\mkG$, 
and the integration is along the line-of-sight (LOS) '$L$' (measured in kpc) from the source to the 
observer. Since the decay of the Faraday RM was already explored for subsonic flows in \citet{SSH06}, 
we limit our analysis here to only transonic flows. Similar to the technique described in \citet{SBS18}, 
we focus on the decaying phase and compute the RM using $512^{2}$ LOS through the simulation 
domain along each of the $x, y,$ and $z$ directions of the simulation box. Unlike incompressible flows, 
we retain $n_{\rm e}$ inside the integral when computing the RM. Since fluctuation dynamo generated 
fields are expected to be statistically isotropic with $\langle {\rm RM} \rangle = 0$ the quantity of 
interest is the normalized standard deviation of RM,  $\sigmarm =  \sigma_{\rm RM}/\sigma_{\rm RM0}$, 
where we define the normalization factor for the intracluster gas as
\EQA
\sigma_{\rm RM0} &=& K\,{\bar{n}_{\rm e}} \frac{b_{\rm rms}}{\sqrt{3}} L \sqrt{\frac{2\pi}{k_{\rm f}L}} 
\sim 250\,{\rm rad\,m^{-2}}\,\left(\frac{\overline{n}_{\rm e}}{10^{-3}\, {\rm cm^{-3}}}\right) \nonumber \\
&&\times \left(\frac{b_{\rm rms}}{2 \mkG}\right)\,
\left(\frac{L}{750\,{\rm Mpc}}\right)^{1/2}\,\left(\frac{l_{\rm f}}{100\,{\rm kpc}}\right)^{1/2}. 
\label{sigmarm}
\ENA 
The above factor derives from a simple model of random magnetic fields, where the equipartition fields 
are assumed to be random with a correlation length $l_{\rm f} = 2\pi/k_{\rm f}$. For example, a fluctuation 
dynamo generated field ordered on the forcing scale is expected to yield $\sigmarm \sim 1$. The final 
value of $\sigmarm$ is taken to be the average of the $\sigmarm$ values obtained from the estimate of 
the standard deviations of ${\rm RM}$ along the three LOS. Furthermore, to explore how $\sigma_{\rm RM}$ 
evolves from steady state as the turbulence decays, we choose to normalize equation~(\ref{sigmarm}) 
using the equipartition value of $b_{\rm rms}$ before turbulent forcing is turned off. 
%%%%%%%%%%%%%%%%%%%%%%%%%%%%%%%%%%%%%%%%%%%%%%%%%%%%%%
\begin{figure}
\begin{center}
\includegraphics[width=1.05\columnwidth]{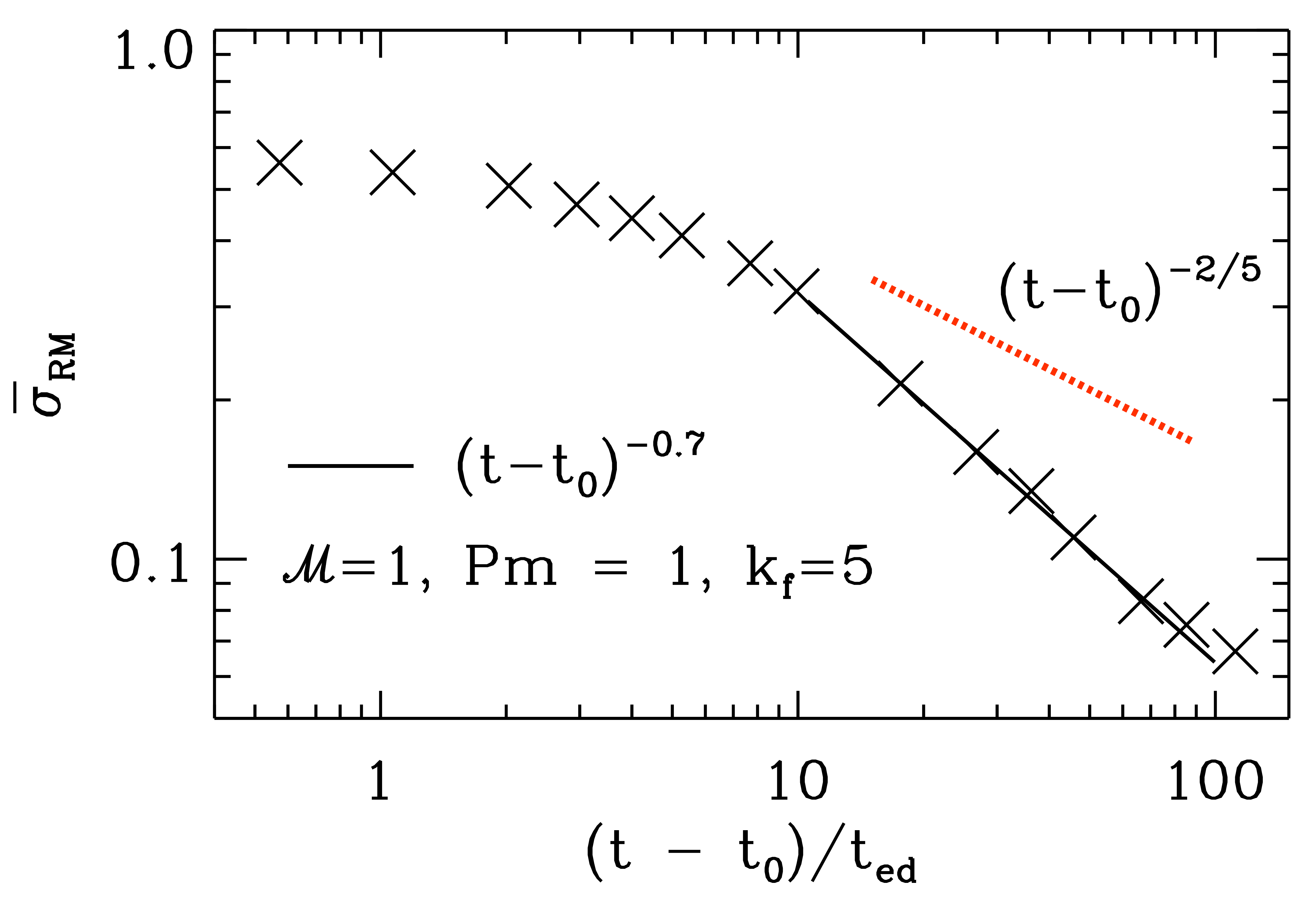}
\vspace{-1em}
\caption{Time evolution of the normalized standard deviation of RM in the decaying phase for the 
transonic run. The red dotted line shows the $t^{-2/5}$ scaling. 
\label{tevol_sigmarm}}
\end{center}
\vspace{-1.5em}
\end{figure}
%%%%%%%%%%%%%%%%%%%%%%%%%%%%%%%%%%%%%%%%%%%%%%%%%%%%%

In the subsonic case \citet{SSH06} found that $b_{\rm rms} \propto u_{\rm rms} \propto t^{-3/5}$ 
and $l_{0} \propto t^{2/5}$, which implies that the observed RM is expected to decrease with time 
as
\EQ
\sigmarm \propto b_{\rm rms}\,l_{0}^{1/2} \propto [(t - t_{0})/t_{\rm ed}]^{-2/5}. 
\EN 
 
From Fig.~\ref{tevol_sigmarm} we find that for the transonic run the steady state value of 
$\sigmarm \approx 0.56$ similar to what was obtained in \citet{SBS18}. This corresponds to a 
$\sigma_{\rm RM} \approx 53\,{\rm rad\,m^{-2}}$ (for $\bar{n}_{\rm e} = 10^{-4}\,{\rm cm^{-3}}, 
b_{\rm rms} = 1\,\mkG$, path length $L = 5\,{\rm Mpc}$, and turbulence forcing scale
$l_{\rm f} = 800\,{\rm kpc}$), in close agreement with the Faraday RM observed in cluster 
outskirts \citep{BCK16}. Thereafter, in the decay phase $\sigmarm$ scales $\approx t^{-5/7}$; 
different from the $t^{-2/5}$ scaling (shown by red dotted line) obtained in the subsonic case.  

\section{Discussion and Conclusions}
\label{conc}

In this paper, we have explored the free decay of turbulence and magnetic fields in the context 
of galaxy clusters. The physical picture that forms the basis of our investigation is as follows. 
We assume that random weak magnetic fields can be amplified to equipartition strengths during 
the epoch of major mergers. Indeed, as shown by \citet{SSH06} seed fields of $10^{-8}{\rm G}$ 
strength can be rapidly amplified to $\mkG$ levels by fluctuation dynamo action assuming that 
the major merger phase lasts for $\approx 3\,{\rm Gyr}$. After this turbulence is expected to decay 
resulting in the decay of the magnetic field. Numerically, these two phases composed of turbulent 
driving and subsequent decay can be simulated using simple idealized simulations. This is the 
approach that we have followed here. In our simulations, weak seed fields are exponentially 
amplified by fluctuation dynamo action and then turbulent driving is turned off once the steady 
state has been captured over sufficient number of eddy turnover times. Keeping in mind that 
cluster turbulence can be both subsonic (near the centre) and transonic (in the outskirts), we 
adjusted the amplitude of the driving such that the resulting flows encompass both the 
above-mentioned regimes. 

Consistent with previous works on fluctuation dynamos the field structure is intermittent in the 
steady state \citep{HBD04, BS05, Cho+09, CR09, BS13, PJR15, Fed16, SBS18}. In the saturated 
state the magnetic spectra exhibit a broad maximum on scales smaller than the forcing scale,  
while the kinetic energy peaks at the forcing scale. Once the forcing is turned off, we find that both 
the rms velocity and the magnetic field decays as a power law in time irrespective of whether the 
turbulent flow in steady state was subsonic or transonic. In cases where the forcing amplitude 
resulted in subsonic flows the power law exponent in the decay phase is close to the $-3/5$ 
scaling reported previously \citep{SSH06}. Analysis of the power spectra in Figs~\ref{spec_sub} 
and \ref{spec_trans} reveals that the decay starts from the high-$k$ end and once the velocity and 
the magnetic field reaches the box scale the decay slows down. 
Interestingly, in the transonic case both $u_{\rm rms}$ and $b_{\rm rms}$ initially decay rapidly 
with time scaling as $t'^{(-1.1)}$ (see Fig.~\ref{tevol_trans_flash}). We found that this is 
accompanied by a decreases of the rms Mach number from $\mathcal{M} \approx 1$ to $\sim 0.1$. 
Beyond this phase, $u_{\rm rms} \sim t^{-0.65}$, while $b_{\rm rms} \sim t^{-5/7}$. 
A naive estimate from the same figure shows that by $10\,t_{\rm ed}$ (after the turbulent driving is 
switched off), the rms magnetic field decreased by a factor $\sim 2.25$, while by $50\,t_{\rm ed}$ it 
has decreased by a factor $\sim 10$. Assuming $l_{\rm f} = 100\,{\rm kpc}$ and 
$u_{\rm rms} = 300\,{\rm km\,s^{-1}}$, the eddy turnover time at the forcing scale, 
$t_{\rm ed} \sim 3\times10^{8}\,{\rm yr} \approx 0.3\,{\rm Gyr}$. Thus, if the saturated field strength 
at the end of the merger event was $\sim 1\,\mkG$, then by $\sim 3\,{\rm Gyr}$ into the decay, 
$b_{\rm rms} \sim 0.5\,\mkG$, which would further decrease to $\sim 0.1\,\mkG$ in about 
$15\,{\rm Gyr}$. While fluctuation dynamo generated fields were shown to possess significant degree 
of coherence even in transonic and supersonic flows \citep{SBS18}, here we applied the analysis 
to compute the evolution of $\sigmarm$ in the decaying phase. In our transonic run, we find that 
$\sigmarm$ also decreases in time as a power law but the decay is $\propto t^{-5/7}$ in comparison 
to the $\sim t^{-2/5}$ scaling obtained for subsonic runs. While the Faraday RM provides information 
on the LOS magnetic field, the intensity and polarization of synchrotron emission will provide 
information on magnetic fields in the sky plane (i.e. perpendicular to the LOS). In a future work, we 
therefore intend to pursue how the synchrotron intensity and degree of polarization evolves in forced 
and decaying phases. 

Apart from galaxy clusters, another intriguing area where our work can be applied concerns magnetic 
fields in elliptical galaxies. Because of lack of radio detection of synchrotron emission from the ISM, 
very little is known about the strength and structure of magnetic fields in these galaxies 
\citep{MB97, MB03}, even though the possibility of fluctuation dynamo action cannot be completely 
ruled out \citep{MS96}. If we assume elliptical galaxies result from mergers of spirals \citep{B92, 
NB03, BJC05, BJC07, NO17}, the parent spirals may already harbour equipartition strength random 
magnetic fields resulting from fluctuation dynamo action. Although turbulence is expected to be initially 
supersonic, it will eventually decay to subsonic levels as these galaxies gradually exhaust their 
molecular and atomic clouds in earlier star formations. For simplicity, let us assume that equipartition 
fields are of strengths $\sim 10\,\mkG$ and that $l_{\rm f} = 100\,{\rm pc}$ and 
$u_{\rm rms} \sim 10\,{\rm km\,s^{-1}}$ implying $t_{\rm ed} \sim 10^{7}\,{\rm yr}$ in the parent spirals. 
Then our estimates from Fig.~\ref{tevol_trans_flash} suggest that by $100\,t_{\rm ed} \sim 1\,{\rm Gyr}$, 
the equipartition field strength would be reduced to $\sim 0.5\mkG$ in elliptical galaxies. In addition to 
random magnetic fields, spirals also possess large-scale mean magnetic fields ordered on kpc scales. 
The large scale and the random magnetic fields are generated by different but related physical 
mechanisms. In the absence of sustained turbulent driving, these helical fields are also expected to 
decay. While this has been studied for non-helical, subsonic turbulent driving \citep[e.g.,][]{BBS14,
B+19}, a systematic analysis of free decay of helical magnetic fields in transonic and supersonic 
turbulence still remains to be performed. 

\section*{Acknowledgements}
I acknowledge computing time awarded at CDAC National Param supercomputing facility, India, 
under the grant "Hydromagnetic-Turbulence-PR" and the use of 'Nova' cluster at IIA. I also thank 
the Science and Engineering Research Board (SERB) of the Department of Science \& Technology 
(DST), Government of India, for support through research grant ECR/2017/001535. 
It is a pleasure to thank Kandaswamy Subramanian for insightful discussions on decaying (M)HD 
turbulence and for his comments on the manuscript. I also thank the anonymous referee 
for a timely and constructive report that helped to improve the quality of the paper.
The software used in this work was in part developed by the DOE NNSA-ASC OASCR Flash 
Center at the University of Chicago.

\bibliographystyle{mnras}

\label{lastpage}

\end{document}